\documentclass[preprint]{revtex4-1}
\usepackage{graphicx}

\usepackage{natbib}

\newcommand{\be}{\begin{equation}}
\newcommand{\ee}{\end{equation}}
\newcommand{\nn}{\mbox{} \nonumber \\ \mbox{} }
\newcommand{\ba}{\begin{eqnarray}}
\newcommand{\ea}{\end{eqnarray}}

\newcommand{\Alfven}{Alfv\'{e}n\,}

\newcommand{\sech}{{\rm \,sech}}

\newcommand\eg{\textit{e.g.}}
\newcommand\cf{\textit{cf.}}

\newcommand{\Bf}{{magnetic field}}

\begin{document}


\title{Relativistic  magnetohydrodynamics  in one dimension}
\begin{abstract}
We derive  a number of  solution for one-dimensional dynamics of relativistic  magnetized plasma that can be used as  benchmark estimates in   relativistic hydrodynamic and magnetohydrodynamic  numerical codes.

First,  we analyze the properties of  simple waves of fast modes propagating orthogonally to the magnetic field in relativistically hot plasma.  The  magnetic and kinetic pressures obey different equations of state, so that the system behaves as a mixture of gases with different  polytropic indices. 
We find  the self-similar solutions for the expansion of hot strongly magnetized plasma into vacuum. 

Second, we derive linear hodograph and Darboux equations for the relativistic Khalatnikov  potential,  which describe arbitrary  one-dimensional isentropic  relativistic  motion of cold magnetized plasma and find their general  and particular solutions.  The  obtained  hodograph and Darboux  equations are  very powerful:    system of highly non-linear,  relativistic, time dependent equations describing  arbitrary (not necessarily self-similar)  dynamics of highly magnetized plasma  reduces  to  a single {\it linear} differential equation.

\end{abstract}

\author{Maxim Lyutikov and Samuel Hadden\\
Department of Physics, Purdue University, \\
 525 Northwestern Avenue,
West Lafayette, IN
47907-2036 }

\maketitle

\section{Introduction}

Expansion of plasma into vacuum is a basic problem in fluid mechanics that has a wide range of applications from heavy nuclei collisions to astrophysics. In nuclear physics,
Belenkij and Landau \cite{BelenkijLandau}
used the hydrodynamical approach to study multiÐparticle production in heavy ion collisions.
 A head-on collision of two highly relativistic nuclei creates a relativistically-compressed hot layer of quarkÐgluon plasma that expands quasi-one-dimensionally \cite{1995NuPhA.595..346R}. 
On a very different scale,  a  wide variety of astrophysical objects like jets from Active Galactic Nuclei (AGN) \cite{2002luml.conf..381B},
Gamma Ray Burst (GRB) \cite{LyutikovJPh} and pulsar winds  and magnetospheres  of a special type of neutron stars - magnetars  \cite{TD93}- 
may contain relativistic strongly
magnetized plasma, in which the energy density of the magnetic field  
dominates over the matter energy density (including kinetic
and thermal energies), $B^2 \geq \rho c^2,\, P$. During explosions the strongly magnetized plasma created by the central source suddenly  expands into a surrounding  low density medium. In case of magnetar flares, the initial dissipation event creates relativistically hot, strongly magnetized fireball, somewhat analogous to Solar coronal mass ejections \cite{lyut06}; 
the fireball accelerates to relativistic velocities \cite{palmer}. 
In long GRBs, when a hot magnetically dominated jet  reaches the surface of the star it  breaks into low density medium \cite{LyutikovJPh,2008MNRAS.388..551T}. 
Similar dynamics may occur in non-stationary outflows in AGNs as well \cite{2010ApJ...722..197L}. 

In all the above mentioned cases, it is expected that at some distances from the source, the \Bf\ is dominated by the toroidal component, while motion is preferentially radial. Thus, velocity is nearly perpendicular to \Bf\ - this type of motion is sometimes called transverse magnetohydrodynamics.  Equations of transverse MHD reduce to fluid equations, but with a complicated equation of state \cite{LLVIII}.  Qualitatively, such plasma behaves like a mixture of fluids with different adiabatic indices, $\Gamma =2$ for the \Bf, and some  $\Gamma$ for the kinetic pressure, usually taken to be $\Gamma=4/3$. Thus, many results of fluid dynamics, which assume a single adiabatic index, become invalid. 

 Numerical investigation of these  phenomena requires the use of relativistic MHD codes that can handle high  magnetization and high kinetic pressures exceeding the rest mass density.
  Exact, explicit  non-linear solutions of fluid equations, and especially  relativistic MHD equations, are rare. Yet they are important for   benchmark estimates of the overall dynamical behavior in   numerical simulations of relativistic flows and  strongly magnetized outflows in particular. 
 In \S \ref{HOT} we find analytical expressions for self-similar expansion into vacuum of  a hot magnetized plasma, considering both the Newtonian and relativistic cases with arbitrary ratios of kinetic and magnetic pressures to rest mass density. 

In addition, at later times, when the whole initial state of plasma is affected by the expansion,  the  expansion dynamics becomes non-self-similar. A classical related  problem is then the expansion of a slab of finite length.

Arbitrary isentropic one-dimensional motion of a fluid is fully integrable. Mathematically this is achieved by exchanging the independent variables $\{t,x\}$  and dependent variables  $\{\rho, v\}$. As a result a system of two non-linear equations (of mass conservation and Euler's law plus assumed  isentropic equation of state) is reduced  to a single 
  {\it linear} equation for the (Legendre-transformed) Bernoulli (Khalatnikov)  potential \citep[\eg][]{LLVI}. This is always possible if the coefficients in these equations do not depend explicitly on time and coordinate.  In the case of isentropic fluid, the equation of state can be correspondingly inverted.
  This transformation is called the hodograph transformation \cite{LLVI}. In \S  \ref{Darboux-ch} we perform a hodograph transformation for relativistic  cold magnetized plasma and  derive the corresponding hodograph  and Darboux equations.

\section{Relativistic expansion of hot magnetized plasma into vacuum}
\label{HOT}

Consider   a relativistic one-dimensional  flow of hot magnetized  fluid  along $x$ direction, carrying \Bf\ in perpendicular direction (so-called transverse  MHD)
The governing of the 
equations are \citep{LLVI}
\ba &&
\partial_t( \gamma \rho) + \partial_x (\gamma \beta \rho) =0
\nn &&
\partial_t T_{00} + \partial_x T_{0x}=0
\nn &&
\partial_tT_{0x}+  \partial_x T_{xx}=0
\nn &&
T_{00}= \gamma^2 w- P -B^2/2
\nn &&
T_{0x}=  \gamma^2 \beta w
\nn &&
T_{xx}=  \gamma^2 \beta^2 w+P +B^2/2
\nn &&
{\cal E} =\rho + 3P +B^2/2, \, w=\rho + 4P +B^2
\label{main0}
\ea 
where  $B$  is the proper (plasma frame)   \Bf\ divided by $\sqrt{4 \pi}$,  $\beta$ and $\gamma$ are fluid's velocities  in terms of the speed of light and Lorentz factors, ${\cal E} $ is the proper energy density, $\rho$ is proper mass density and  $w$ is proper enthalpy. The adiabatic index for the kinetic pressure is assumed to be $\Gamma=4/3$. 

Let us consider a semi-space $x<0$ occupied by  homogeneous hot plasma with  density $\rho_0$, \Bf\ $B_0$ and pressure $P_0$. The initial state is assumed to be homogeneous.  (At time $t=0$ a barrier located at $x=0$ is removed; plasma starts expanding in the positive $x$ direction, while a rarefaction wave is launched in the negative $x$ direction.
We
introduce two dimensionless parameters describing the initial plasma magnetization and the ratio of kinetic to magnetic pressures:
\ba &&
\sigma _0 = B_0^2/\rho_0
\nn &&
{\cal B}= 2 P_0/B_0^2
\ea
Parameter $\sigma$ \cite{kc84} measures the  importance of relativistic effects of  magnetic fields: for $\sigma \geq 1$ the \Alfven velocity approaches the speed of light. 
Parameter ${\cal B}$ is the conventional  plasma beta parameter:  the ratio of kinetic and magnetic pressures.

The relevant  speed of propagation of disturbances is  the fast  magnetosonic speed $c_f$ \citep{1957PhRv..108.1357H}
\ba &&
c_f^2 =  { 3 B^2 +4 P \over 3( B^2 + 4 P +\rho)}=v_A^2 + c_s^2 = { (3+ 2 {\cal B}) \sigma \over 3(1+ \sigma(1 + 2 {\cal B}))}
\nn &&
v_A^2={\sigma \over 1+ \sigma(1 + 2 {\cal B} )}
\nn &&
c_s^2 =  { 2 {\cal B} \sigma\over 3(  1+\sigma +{\cal B} \sigma)}
\label{cs}
\ea
where $v_A$ is \Alfven velocity and $c_s$ is sound speed.
Note that for $\sigma =1/2$, the fast speed is independent of ${\cal B}$ and  equals $c_f =1/\sqrt{3}$. In this case the contribution of kinetic pressure to total pressure is compensated by its contribution to effective mass density.
 
For transverse  MHD (when the motion is perpendicular to the \Bf), the  induction equation and matter conservation require $B/B_0 = \rho/\rho_0$. 
Assuming that all the quantities depend on the self-similar variable $\eta =x/t$, Eqns. (\ref{main0}) give
\ba &&
\rho_1 ' = {1 -\eta \beta \over\eta - \beta } \gamma^2 \rho_1 \beta'
\label{1111}
\\ && 
\gamma^2 (\eta - 2 \beta +\eta \beta^2) \left( 4 P + \rho_1 ( \rho_0 + B_0^2 \rho_1)\right) \beta ' +
\nn &&   \left( \eta \beta(\rho_0 + 2 B_02^2 \rho_1) - \beta^2 (\rho_0+B_0^2 \rho_1) - B_0^2 \rho_1 \right) \rho_1 ' - (1-4 \eta \beta - 3 \beta^2) P'=0
\ea
where $\rho_1 = \rho/\rho_0$ and primes denote derivative with respect to $\eta$ (note, that if the initial state is not homogeneous, the self-similar solutions may depend on combination $x^\alpha/t$).

Eliminating pressure and \Bf\ in the initial state in favor of $\sigma $ and ${\cal B}$, $P=P_0 (\rho/\rho_0)^{4/3} $,  $P_0 = ({\cal B} \sigma/2) \rho_0$, $B_0 = \sqrt{ \sigma  \rho_0}$, we find
\be
(\eta -\beta)^2 - {1-\eta^2 \over \gamma^2} \sigma \rho_1 - {2\over 3} \left(1+4  \beta \eta - 3 \eta^2 + \beta^2 ( \eta^2-3) \right) {\cal B} \sigma \rho_1^{1/3} =0
\label{main1}
\ee
Changing to Doppler factors
\ba && 
\delta _\beta = \sqrt{ 1+\beta \over 1-\beta}
\nn &&
\delta _\eta = \sqrt{ 1+ \eta \over 1-\eta},
\label{delta1}
\ea
Eqns. (\ref{1111}-\ref{main1})  become
\ba &&
(\delta _\beta^2 - \delta _\eta^2) ^2 - 4  \delta _\beta^2  \delta _\eta^2\sigma \rho_1 +
{4 \over 3} \left( \delta _\beta^4 - 4  \delta _\beta^2 \delta _\eta^2 +  \delta _\eta^4 \right){\cal B} \sigma \rho_1^{1/3} =0
\label{1}
\\ &&
(\delta _\beta^2 + \delta _\eta^2)\rho_1 {\partial \delta _\beta \over \partial \delta _\eta} + \delta _\beta (\delta _\beta^2 - \delta _\eta^2) 
{\partial\rho_1 \over \partial \delta _\eta} =0
\label{2}
\ea

It is  more convenient  to change to  a new variable
$U = \sqrt{ \rho_1 \sigma} $ and redefine parameter $ {\cal B}_1= (2/3) {\cal B} \sigma ^{2/3}$ (in this case the resulting equations (\ref{11}-\ref{12}) depend only on one parameter ${\cal B}_1$; also, in the cold limit $U$ becomes a four-velocity    of \Alfven waves    $U = \beta_A/\sqrt{1-\beta_A^2}$ \cite{Simplewaves}).
In terms of variables $\beta - U$, Eqns (\ref{1}-\ref{2}) become
\ba && 
(\delta _\beta^2 - \delta _\eta^2) ^2 - 4  \delta _\beta^2  \delta _\eta^2 U^2 + 2 {\cal B}_1 U^{2/3}  \left( \delta _\beta^4 - 4  \delta _\beta^2 \delta _\eta^2 +  \delta _\eta^4 \right)=0
\label{11} 
\\  && 
(\delta _\beta^2 + \delta _\eta^2) U  {\partial \delta _\beta \over \partial \delta _\eta}- 2 \delta _\beta (\delta _\beta^2 - \delta _\eta^2) 
{\partial U \over \partial \delta _\eta}=0
 \label{12} 
\ea

Previously, Lyutikov \cite{Simplewaves} found  simple analytical solution of  Eqns. (\ref{11}-\ref{12})  for  simple waves in cold magnetized plasma expanding into vacuum.
If initially the plasma is at rest,  and the  \Alfven velocity in the unperturbed medium  $v_{A,0}$ is given by  corresponding Doppler factor $\delta_{A,0}= \sqrt{(1+v_{A,0})/(1-v_{A,0})}$, in the limit
$ {\cal B}_1=0$,  Eqns. (\ref{11}-{12})  have solutions
\be 
\delta_\beta = \delta_\eta^{2/3} \delta_{A,0}^{\pm2/3}
, \,
\delta_A ={ \delta_{A,0}^{2/3}  \delta_\eta^{\mp1/3}}
\label{EE}
\ee

In case of hot plasma,  $ {\cal B}_1\neq 0$,
Eq. (\ref{11}) can be resolved for $\delta_\beta$
\be
{ \delta^2 _\beta \over \delta _\eta^2} = {1+4  {\cal B}_1 U^{2/3} +  2 U^2 \pm  2  U^{1/3}  \sqrt{ 3{\cal B}_1^2 U^{2/3} + U^{4/3} + U^{10/3} + { \cal B}_1(1+4 U^2)}  \over 1+ 2   {\cal B}_1 U^{2/3} }  \equiv f(U). 
\label{f}
\ee
This gives a general solution for the Doppler factor in terms of initial parameters $ {\cal B}_1= (2/3) {\cal B} \sigma ^{2/3}$  and local density $U = \sqrt{ \rho_1 \sigma} $.

Eq.  (\ref{12}) then becomes
\be
{\partial U \over \partial \ln \delta _\eta} = { f (1+f^2) U \over 2 f( 1-f^2) - (1+f^2) U \partial_ U  f}
\ee
or, using  the explicit form of $f$, Eq. (\ref{f}), 
\be
{\partial \ln \delta _\eta \over \partial U } = - {  12  {\cal B}_1^2 U^{2/3} + 9 U^{4/3} +  {\cal B}_1 (7+16 U^2)  \over 3 (1+2  {\cal B}_1 U^{2/3} U^{2/3}
 \sqrt{( {\cal B}_1+U^{4/3})( 1+ 3  {\cal B}_1 U^{2/3}+  U^2)}  }  
 \label{Sol1}
\ee
Changing independent variables $\delta _\eta \rightarrow U$, we find:
\be
\ln \delta _\eta = - \int ^U { 12  {\cal B}_1^2 y^{2/3} + 9 y^{4/3} +  {\cal B}_1 (7+16 y^2)
 \over 3 (1+2  {\cal B}_1 y^{2/3} ) y^{2/3}
 \sqrt{( {\cal B}_1+y^{4/3})( 1+ 3  {\cal B}_1 y^{2/3}+  y^2)} } dy
  \label{Sol2}
 \ee
Eq. (\ref{Sol2}) gives a general solution for  simple  fast waves in relativistic magnetized fluid with arbitrary ratios of magnetic and kinetic pressure to rest mass, with kinetic part of the pressure obeying adiabatic law with $\Gamma =4/3$.   Eq. (\ref{Sol2}) expresses implicitly  the density $U = \sqrt{ \rho_1 \sigma} $ in terms of the  self-similar variable $\eta = x/t$. 

\begin{figure}[h!]
 \begin{center}
\includegraphics[width=.49\linewidth]{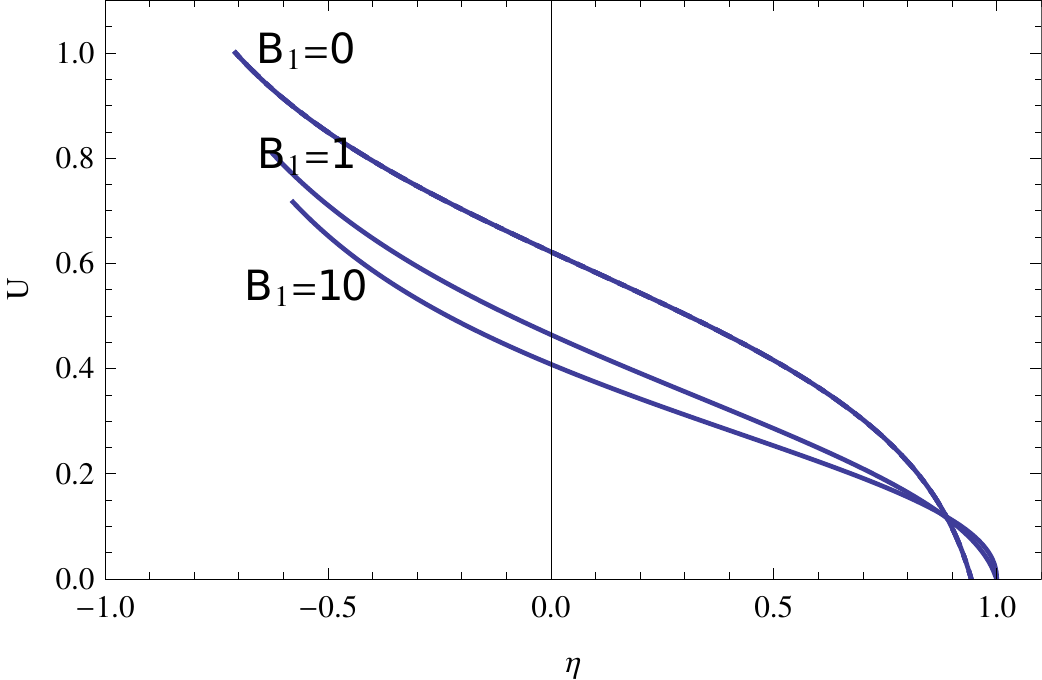}
\includegraphics[width=.49\linewidth]{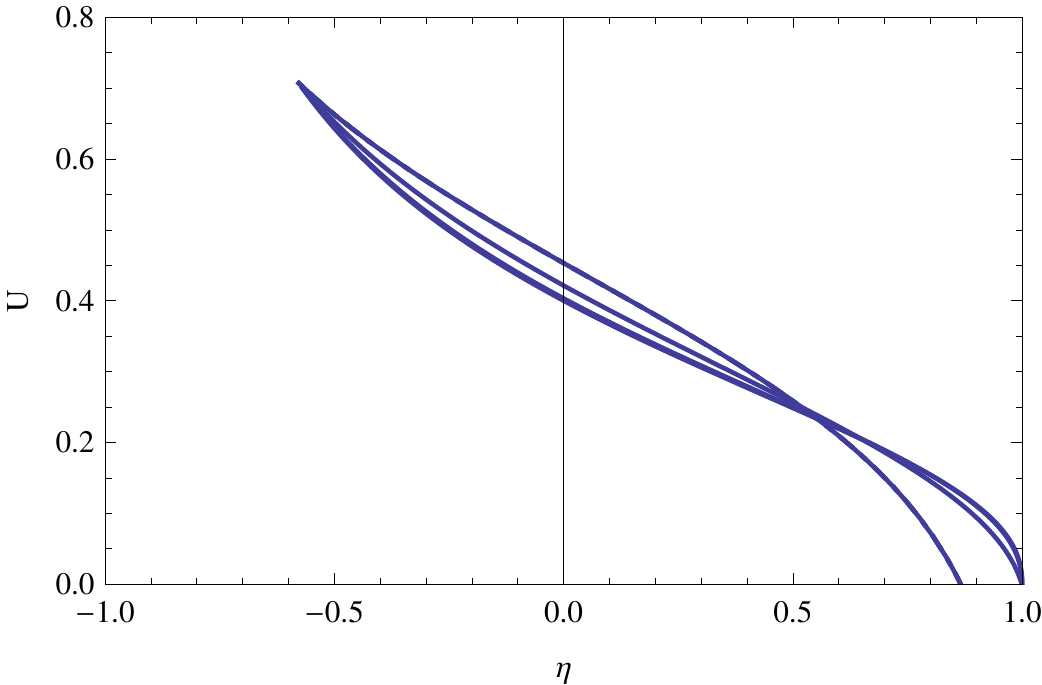}
\end{center}
\caption{ Four-velocity of the relativistic self-similar expansion for different values of $ {\cal B}_1= (2/3) {\cal B} \sigma ^{2/3}$,  ${\cal B}= 2 P_0/B_0^2$ . Plotted is the value of  $U = \sqrt{ \rho_1 \sigma} $ as a function of the  self-similar parameter for various values of the kinetic pressure parameter $ {\cal B}_1$ and magnetization. {\it Left Panel}:  $\sigma =1$, {\it Right Panel}  $\sigma =1/2$.  The upper curve, corresponding to   $ {\cal B}_1=0$, is, in fact  {\it two} coincident curves, plotted using the known explicit solution for relativistic expansion of magnetized cold gas into vacuum, Eq. (\ref{EE}), see also \cite{Simplewaves}, and the implicit solutions (\ref{Sol2}) for zero kinetic pressure. In this case the front of the rarefaction wave is located at $\eta = - \sqrt{ \sigma /(1+\sigma)}$. For $\sigma =1/2$, the front of the rarefaction wave is located at $\eta = - \sqrt{ \sigma /(1+\sigma)}= -  1/\sqrt{3} $}
\label{Uofeta}
\end{figure}

\begin{figure}[h!]
 \begin{center}
\includegraphics[width=.49\linewidth]{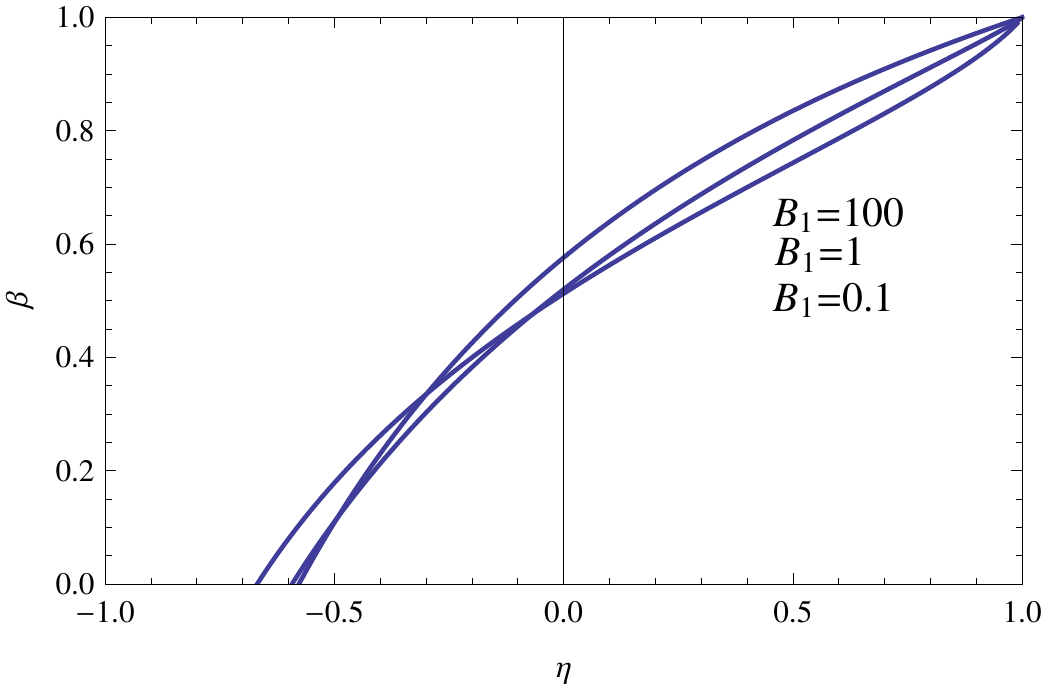}
\includegraphics[width=.49\linewidth]{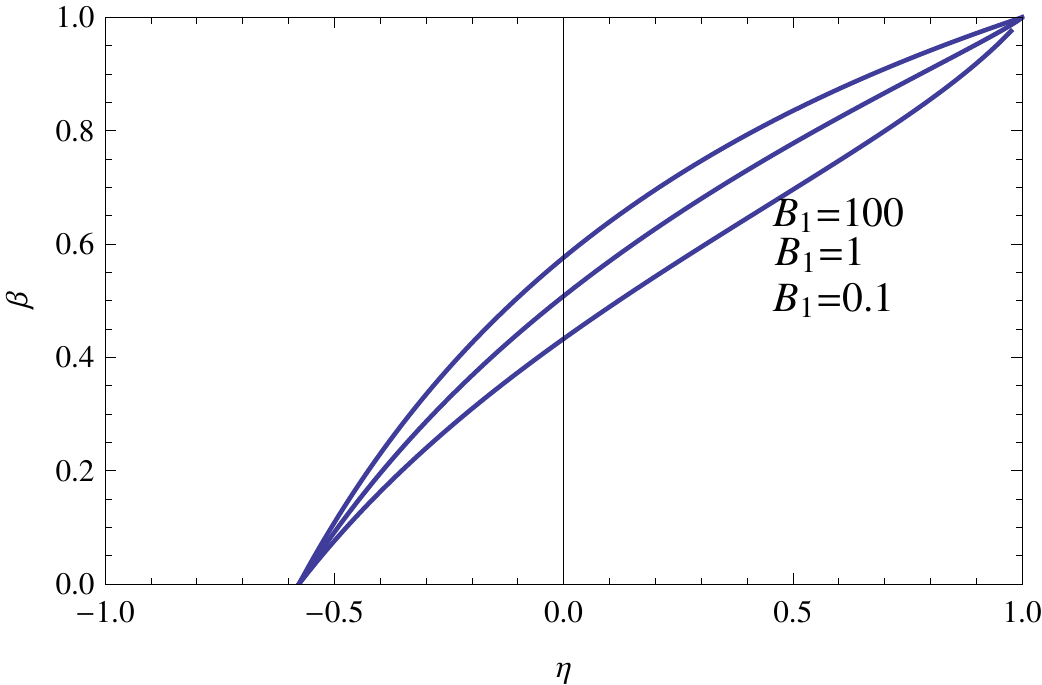}
\end{center}
\caption{Velocities  for relativistic expansion of hot plasma into vacuum $\sigma =1$ (Left Panel), $\sigma =1/2$ (Right Panel).}
\label{betaofeta}
\end{figure}

The lower limit of integration in Eq. (\ref{Sol2}) can be found from the conditions on the front characteristics propagating into undisturbed plasma. 
The rarefaction wave propagates into the undisturbed medium with the local fast velocity  (see Eq. (\ref{cs})):
\be
c_{f,0}^2 = { {\cal B}_1 \sigma^{1/3} + \sigma \over 1 + 3  {\cal B}_1 \sigma^{1/3} +\sigma}
\ee
This corresponds to
\ba &&
\delta_ {\eta _0}   = \sqrt { 1-c_{f,0} \over 1+ c_{f,0}} 
\nn &&
U_0 = {1\over 2} \left( {1\over \delta_ {\eta _0}   } -  \delta_ {\eta _0}  \right)
\ea
(note the  signs of velocities  in the  definition of  $\delta_ {\eta _0} $: the rarefaction wave propagates in the direction opposite to the flow).
 At the front of the rarefaction wave
 $\delta_ {\eta} = \delta_ {\eta _0}  $. Thus, a particular solution corresponding to expansion into vacuum of a medium initially at rest is
 \be
\ln { \delta _\eta \over \delta_ {\eta _0} }  = - \int_{U_0 } ^U { 3 (1+2  {\cal B}_1 y^{2/3} y^{2/3}
 \sqrt{( {\cal B}_1+y^{4/3})( 1+ 3  {\cal B}_1 y^{2/3}+  y^2)} \over 12  {\cal B}_1^2 y^{2/3} + 9 y^{4/3} +  {\cal B}_1 (7+16 y^2)} dy
  \label{Sol3}
 \ee
 Eq. (\ref{Sol3}) gives a solution for relativistic expansion of hot  magnetized plasma into vacuum, see Fig. \ref{Uofeta}. Corresponding velocities are plotted in  Fig. \ref{betaofeta}.

One can verify that  Eq. (\ref{Sol3})  reproduces the known result for cold magnetized plasma \cite{Simplewaves}.
For  zero kinetic pressure, ${\cal B}_1=0$,  we find
\ba &&
f(U)= \delta _U ^{\pm 1}
\nn &&
 \delta_{U}= \delta_{U_0}^{2/3}   \delta_\eta^{\pm 1/3}
 \label{GG}
\ea
in agreement with \cite{Simplewaves}. 

For zero  kinetic pressure, $ {\cal B}_1 \rightarrow 0 $,  relations (\ref{Sol3}-\ref{GG}) imply that  the front of the rarefaction wave is located at $\eta_{RW} = - \sqrt{\sigma /(1+\sigma)}$, while 
for large kinetic pressure, $ {\cal B}_1 \rightarrow \infty$,  the front of the rarefaction wave is located at $\eta_{RW}  =  - 1/\sqrt{3}$.
The vacuum interface in the limit  $ {\cal B}_1 \rightarrow \infty$ approaches the  speed of light.

 For a fixed values of $\sigma> 1/2$, the  increase of kinetic pressure leads to larger values of negative  $\eta_{RW}$  (smaller absolute values of   $\eta_{RW}$) due to effective increase in plasma inertia. For $\sigma =1/2$, the front of the rarefaction wave is located at fixed $\eta = -  1/\sqrt{3}$ for any $ {\cal B}_1$.
Finally, we note that the Riemann invariants in a transverse relativistic MHD,
does not have a representation in simple functions.

\section{Self-similar expansion of unmagnetized relativistic  fluid}

In this section we simplify the above relations for the case of simple waves in relativistic unmagnetized fluid. Adopting a  polytropic EoS,  we will calculate simple waves in plasma with adiabatic index of $4/3$, yet for a finite ratio $P/\rho$, without assuming that the speed of sound equals $1/\sqrt{3}$. 
This is an important  step since most codes use a single value of adiabatic index, but for finite ratios  $P/\rho$.

For a finite ratio of $P/\rho$, the
 speed of sound equals
\be
c_s^2 = {4 P_0 \over 3 (4 P_0 +\rho_0)},
\ee
while Eqns. (\ref{main1}) become
\ba &&
(\eta -\beta)^2 -  {4 P_0 \over 3 \rho_0}  \left( 1- 3\eta^2 + 4 \eta \beta -(3-\eta^2) \beta^2  \right) \rho_1^{1/3} =0
\label{21}
\\ &&
\rho_1 ' = {1 -\eta \beta \over\eta - \beta } \gamma^2 \rho_1 \beta'
\label{22}
\ea

In case of negligible density $\rho_0 \rightarrow 0$, this immediately gives the ultra-relativistic limit $c_s =1/\sqrt{3}$,
\ba && 
\beta = { 2 \eta \pm \sqrt{3} (1-\eta^2)\over 3 -\eta^2}
\nn &&
\delta_\beta = { \delta _\eta \over \sqrt{ 2 -\sqrt{3}}} 
\label{inf}
\ea
The front  of the rarefaction wave is located at $\eta_{RW} = -1/\sqrt{3}$, while on the vacuum side expansion proceeds with the speed of light. 

In the ultra-relativistic limit the flow lines $dx/dt= \beta$ are given by
\be
{t\over t_0} = {1\over \sqrt{1-\eta^2}} \left({1+\eta \over 1-\eta}  \right)^{\sqrt{3}/2}
\ee
The characteristics satisfy $dx/dt= (\beta+c_s)/(1+\beta c_s)$, which gives
\be
{t\over t_0} = {1\over \sqrt{1-\eta^2}} \left({1+\eta \over 1-\eta}  \right)^{1/\sqrt{3}}
\ee
(the other  characteristics are  straight lines).

\begin{figure}[h!]
 \begin{center}
\includegraphics[width=.99\linewidth]{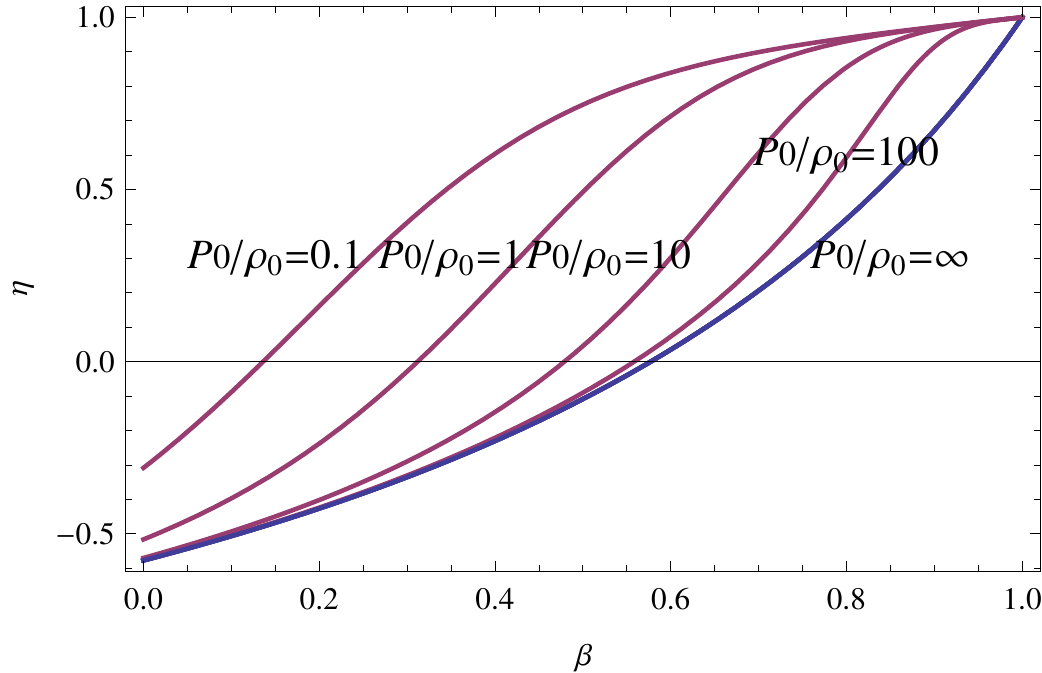}
\end{center}
\caption{Self-similar expansion of  relativistic unmagnetized fluid with adiabatic index $\Gamma =4/3$. Top to bottom $P_0/\rho_0= 0.1, \, 1,\, 10,\, 100$. The curve $P_0 /\rho_0 =\infty$ is given by  Eq. (\ref{inf}).   The  vacuum interface propagates with the  speed of light.  In the ultra-relativistic limit, $P_0/\rho_0
\rightarrow \infty$ the  front of the rarefaction wave is located at $\eta_{RW} = -1/\sqrt{3}$.  }
\label{betalimitofeta3}
\end{figure}

In case of finite density, 
  introducing Doppler factors, Eqns. (\ref{21}-\ref{22}) give
\ba && 
\rho_1 ={3^{1/3} \over 2} \left( {(\delta_\eta^2 -\delta_\beta^2)^2 \over 4 \delta_\beta^2 \delta_\eta^2 - \delta_\beta^4 - \delta_\eta^4}   {\rho_0 \over P_0}\right)^{1/3}
\label{rrr}
\\ &&
\left( 3 \delta _\eta^4 \delta_\beta - 16 \delta_\eta^2 \delta_\beta^3 + 3 \delta_\beta^5  \right)\delta_\beta ' + 4 \delta_\eta \delta_\beta^4=0
\label{rr1}
\ea
Which can be integrated 
\be
 \delta _\eta^2 = { \sqrt{3} \delta_\beta^2 ( C+ \delta_\beta^{3 \sqrt{3}}) \over
 (3+2 \sqrt{3}) C - (3-2 \sqrt{3})\delta_\beta^{3 \sqrt{3}} }
 \label{eta}
 \ee 
 The  constant of integration  $C$ in Eq. (\ref{eta}) 
  can be found from the condition that at the front of the rarefaction wave, propagating with velocity $-c_{s,0}$ and located at  $  \eta_0 = -c_{s,0}, \delta_{\eta,0}=\sqrt{(1+\eta_0)/(1-\eta_0)}  $ the fluid is at rest, $\beta=0, \delta_\beta =1$. 
This gives
\be
C= - { \sqrt{3} + (3-2\sqrt{3})  \delta_{\eta,0}^2 \over \sqrt{3} - (3+2\sqrt{3})  \delta_{\eta,0}^2} 
\label{C}
\ee
Equations (\ref{rrr}, \ref{eta}, \ref{C}) give an analytical  solution to the problem of self-similar expansion of fluid with $\Gamma =4/3$ into vacuum, valid for arbitrary ratios of kinetic  pressure to mass density, see Fig. \ref{betalimitofeta3}.

\section{Non-similar  expansion of cold magnetized plasma: relativistic hodograph and Darboux equations}
\label{Darboux-ch}


Relativistic hodograph transformation was derived by  Belenkij \& Landau \cite{BelenkijLandau} for polytropic fluid. Here we first re-derive the corresponding equation for cold  magnetized plasma and then transform it to a normal form, where Riemann invariants are taken as independent variables. In case of regular fluids the internal energy and the corresponding  hodograph equations are defined in term of temperature, which is zero in the case of cold magnetized fluid. As we will see below, the role of temperature is taken by the proper enthalpy.


The relativistic hodograph transformation is achieved by introducing the Khalatnikov potential $\phi$ \citep{LLVI}
\ba &&
 \gamma \beta \tilde{w } = \partial_x \phi 
\nn && 
 \gamma \tilde{w }  = - \partial_ t \phi
 \nn && 
 \tilde{w } = {w \over \rho}
 \ea
 In the non-relativistic limit the corresponding equations are the  condition on potential flow and the  Bernoulli equation. Below we drop the tilde sign over the enthalpy: $w$ is then  the enthalpy per unit mass:
 \be
 w= { \rho+ B^2 \over \rho}  = {1 \over 1- \beta_A^2}
 \ee
 
 The differential of the  Khalatnikov potential is 
 \be
 d \phi= \partial_ x \phi dx +  \partial_ t \phi dt =  \gamma \beta w dx -  \gamma w dt
 \ee
 Next we employ Legendre transform of $\phi$ with respect to variables $\{x,t\}$. The transformed potential $\chi$ becomes
 \ba &&
 \chi =  \phi -   \gamma \beta w x +   \gamma w t
 \nn &&
 d  \chi = (t - \beta x) \gamma dw -(x- \beta t) w \gamma^3 d \beta
 \ea
 Variables $t$ and $x$ are then expressed from the potential $\chi$ as
 \ba &&
 t= \gamma {\partial \chi \over \partial w} - { \beta \over \gamma w}   {\partial \chi \over \partial \beta}
 \nn &&
 x= \beta t -  { 1 \over  \gamma^3 w }    {\partial \chi \over \partial \beta}
 \ea
 
 These equations can be modified if we introduce rapidity $\beta = \tanh r$:
 \ba &&
 t= \cosh r  {\partial \chi \over \partial w} -{\sinh r \over w}  {\partial \chi \over \partial r}
 \nn &&
 x= \sinh r   {\partial \chi \over \partial w} -{\cosh r \over w}  {\partial \chi \over \partial r}
 \ea
 Comparing with  Ref. 
\cite{BelenkijLandau}    Eq. (4.12), we see that in cold magnetized plasma proper enthalpy $w$ plays a role of a temperature, while 
the speed of sound $c_s$ -- \Alfven velocity -  is given by
\be
c_s^2 = v_A^2 = {\partial P \over  \partial {\cal E}} = {B^2 \over B^2 +\rho} = 1-1/w
\ee

The relativistic  hodograph equation is then obtained from the continuity equation by transforming to independent variables $r $  and $w$. 
\be
\partial^2 _r \chi  - w \partial _w\chi + (1-w) w \partial^2 _w \chi =0
\label{hodog}
\ee
This is relativistic hodgoraph equation for one-dimensional motion of cold magnetized plasma. 
Eq. (\ref{hodog}) reduces to the one obtained by Belenkij and Landau for hot fluid with a substitution $w= (\rho + B^2 ) /\rho \rightarrow T$.
A general  separable solution of Eq. (\ref{hodog}) is
\be
\chi=  e^{ C r}  \, w\,  {_2 F_1} ( 1-C, 1+C,2,w)
\ee
where, we remind, $r$ is the rapidity, $\beta = \tanh r$, and ${_2 F_1} $ is the hypergeometric function.

Let us transform the hodgoraph equation (\ref{hodog}) taking  the Riemann invariants as independent variables.
The Riemann invariants are  \citep{Simplewaves}
\ba && 
J_1 = \log \delta_A^2 \delta_\beta
\nn &&
J_2= \log {\delta_A^2  \over \delta_\beta}
\label{46}
\ea
where
\ba && 
 \delta_A = \sqrt{ 1+ \beta_A \over 1-\beta_A} = \sqrt{ 1+ \sqrt{1-1/w} \over 1 - \sqrt{1-1/w}}
 \nn &&
  \delta_\beta=  \sqrt{ 1+ \beta \over 1-\beta} = \sqrt{ 1+ \tanh r \over 1-\tanh r} 
  \label{delta}
  \ea
  are corresponding Doppler factors. 

Using Eqns (\ref{delta}), the Riemann invariants become
\ba &&
J_1 =  2 {\rm arctanh} \sqrt{1-1/w} +r
\nn &&
J_2 =  2 {\rm arctanh}   \sqrt{1-1/w} -r
\ea

Thus, the rapidity $r$ and the proper enthalpy are
\ba &&
r= {J_1 - J_2 \over 2}
\nn &&
w=  \cosh ^2 {J_1+J_2 \over 4}
\ea

Changing the independent variables in the hodograph equation (\ref{hodog}), we derive the relativistic Darboux equation for the Khalatnikov potential as a function of the Riemann invariants:
\be
\partial _{J_1} \partial _{J_2} \chi + {1\over 4} { \partial _{J_1} \chi  +  \partial _{J_2} \chi  \over \sinh {J_1+J_2 \over 2} }=0
\label{DarbouxR}
\ee
In the non-relativistic limit $ J_1 \rightarrow \beta+ 2 \beta_A, \, J_2 \rightarrow 2 \beta_A- \beta$, $\sinh {(J_1+J_2 )/ 2 } \approx {(J_1+J_2)/ 2} $ and Eq. (\ref{DarbouxR}) reduces to the  non-relativistic Darboux equation for a fluid with adiabatic index of $\Gamma=2 $, Eq. (\ref{Darboux}) 

Using constancy of the Riemann invariants on the characteristics, one finds (\cf, Eq, (\ref{xtr}))
\ba &&
\partial_{J_2} x = \tanh \left( {  J_1 - 3 J_2 \over 4} \right) \partial_{J_2} t
\nn &&
\partial_{J_2} x = \tanh \left( { 3 J_1 - J_2 \over 4} \right)\partial_{J_2} t
\label{xtr}
\ea
System (\ref{xtr}) can be written as a single equation for time  
\be
\partial_{J_1} \partial_{J_2} t  +{3\over 4}  { \sech^2  \left( {  J_1 - 3 J_2 \over 4} \right) \partial_{J_1}t +\sech^2   \left( { 3 J_1 - J_2 \over 4} \right)  \partial_{J_2} t \over \tanh \left( { 3 J_1 - J_2 \over 4} \right) - \tanh \left( {  J_1 - 3 J_2 \over 4} \right)}=0
\label{DDD}
\ee
This is the relativistic analogue of the Darboux equation for the  time variable. Solutions of Eqs.  (\ref{DDD}-\ref{xtr})   give time and spacial coordinate as   functions of two Riemann invariants. These solutions can then be inverted for $J_{1,2} (x,t)$.

\section{1D expansion of a slab of gas: solutions of the hodograph equation}

Consider a slab of gas initially at rest, occupying region $0<x<L$ and expanding into vacuum $x>L$. At $x=0$ there is an impenetrable wall, Fig. \ref{fig:relplots}.
In the initial state  the Riemann invariants are $J_{1,0}=J_{2,0}= 2 c_{s,0}$ where $ c_{s,0}=v_{A,0} $ is the sound (\Alfven) velocity in the initial state.
The boundary conditions  for this problem are \citep[][problem after parag. 105]{LLVI}: zero velocity at  the wall and constancy of the first Riemann invariant on the characteristics that leaves the wall at the moment of reflection:
\ba && 
\left. 
{\partial \chi \over \partial \beta} \right|_{\beta =0 } = 0
\nn  && 
\chi(J_1 = J_{1,0}) = 0
\label{boundary}
\ea

Numerical solutions for Alfv{\' e}n and sound speeds are found and plotted in both the non-relativistic (as discussed in Appendix \ref{non-relattt}) and relativistic
case.  
In both cases finding a numerical solution involves inverting a pair of functions of
the Riemann invariants, \(\left\{t\left(J_1,J_2\right) ,x\left(J_1,J_2\right)\right\}\), to obtain functions of position and time giving the Riemann
invariants, \(\left\{J_1(x,t),J_2(x,t)\right\}\). In the non-relativistic case the initial functions  \(\left\{t\left(J_1,J_2\right) ,x\left(J_1,J_2\right)\right\}\)
are functions represented by Eq.{'}s (\ref{t}) and (\ref{xx}) whereas in the relativistic case these functions are determined using numerical methods.  The
relevant quantities, Alfv{\' e}n and sound speed, can then be expressed as functions of position and time.  The numerical solutions in both relativistic
and non-relativistic cases involve separate calculations for the regions of expansion affected by the reflection of the initial rarefaction wave
at the position of the wall, \(x=0\), and the regions unaffected by this reflection (as well as the boundary between these regions). The former
region will be referred to as the {``}complex{''} region and latter as the {``}simple{''} region.

\subsection*{B. Numerical Solutions of Relativistic Expansion}

Initial conditions in the relativistic case are given, by initial Alfv{\' e}n speed \(\beta _A(x,0)=\tanh (1),\text{}\)initial velocity
\(\beta (x,0)=0\), the plasma occupies the region \(0<x<\tanh (1)\) with an impenetrable wall at \(x=0\) and the plasma expands into vacuum at
\(t=0\).  The initial conditions are such that reflection of the initial
rarefaction wave occurs at \(t=1\). The Initial Alfv{\' e}n speed, \(\tanh (1)\simeq .762\), was selected as a speed somewhat near the speed
of light that gives convenient initial values for the Riemann invariants: \(2=J_{1,0}=J_{2,0}\). 

In the case of relativistic expansion, no simple analogue to Eq. (\ref{t}) that satisfies Eq. (\ref{DDD}) and gives time as a function of the Riemann
invariants in the complex wave region is readily available. Therefore, in the complex region, time must be numerically calculated as a function
of the two Riemann invariants, \(J_{1,2}\). The first step in this calculation is accomplished by numerically solving for a function satisfying
the adjoint differential operator corresponding to Eq. (73) subject to boundary conditions as outlined in \citep[][Ch. 5, Sec. 2, (Eq.'s
3 and 3')]{1953mmp..book.....C}, obtaining a solution \(R\left[J_1,J_2; \alpha ,\beta \right]\).
 In the present calculations this function is solved numerically  after specifying boundary values based on $\alpha $ and $\beta $. 
Exploiting the symmetry of the Riemann's function, \(R\left[J_1,J_2; \alpha ,\beta \right]\)  \citep[][Ch. 5, Sec. 2, (Eq.'s
3 and 3')]{1953mmp..book.....C},, one obtains an expression for time
given as a function of \(J_1 \text{ and } J_2\) satisfying (\ref{DDD}) : 
\be
t\left[J_1,J_2\right]=R\left[2\tanh ^{-1}\left(\beta _{A,0}\right),2\tanh ^{-1}\left(\beta _{A,0}\right);J_1,J_2\right]
\ee
The initial value of both Riemann invariants, \(2\tanh ^{-1}\left(\beta _{A,0}\right)\), corresponds to the initial conditions of \(\beta _{t=t_0}=0\)
and \(\beta _{A,t=t_0}=\)\(\beta _{A,0}\). Position can then be calculated as a function of the Riemann invariants by integrating by parts Eq.
(\ref{xtr}).

\begin{figure}[h!]
 \begin{center}
\includegraphics[width=.49\linewidth]{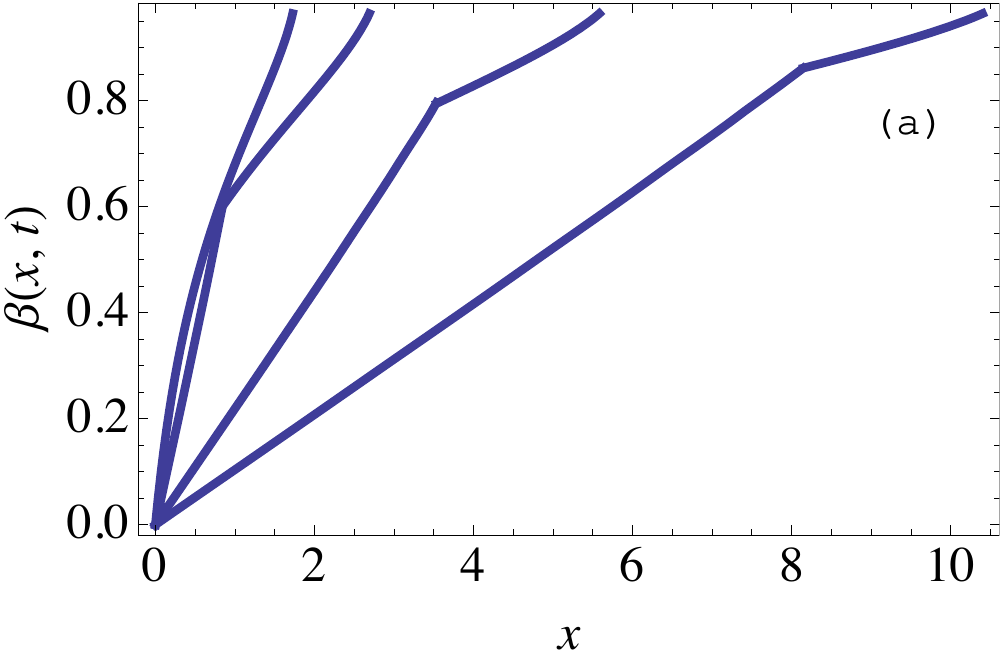}
\includegraphics[width=.49\linewidth]{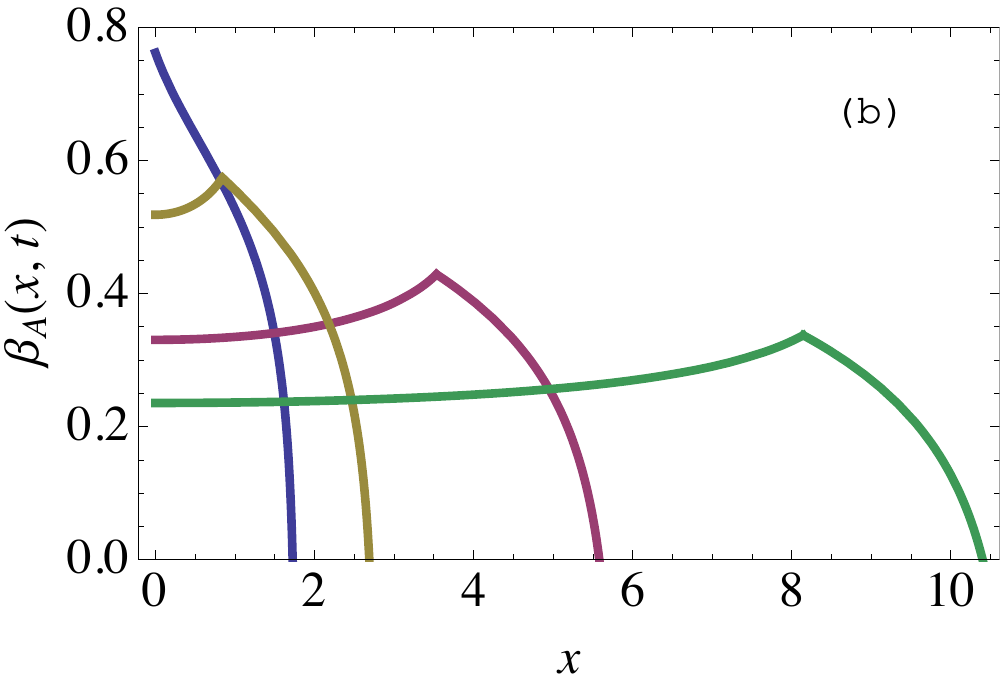}
\includegraphics[width=.49\linewidth]{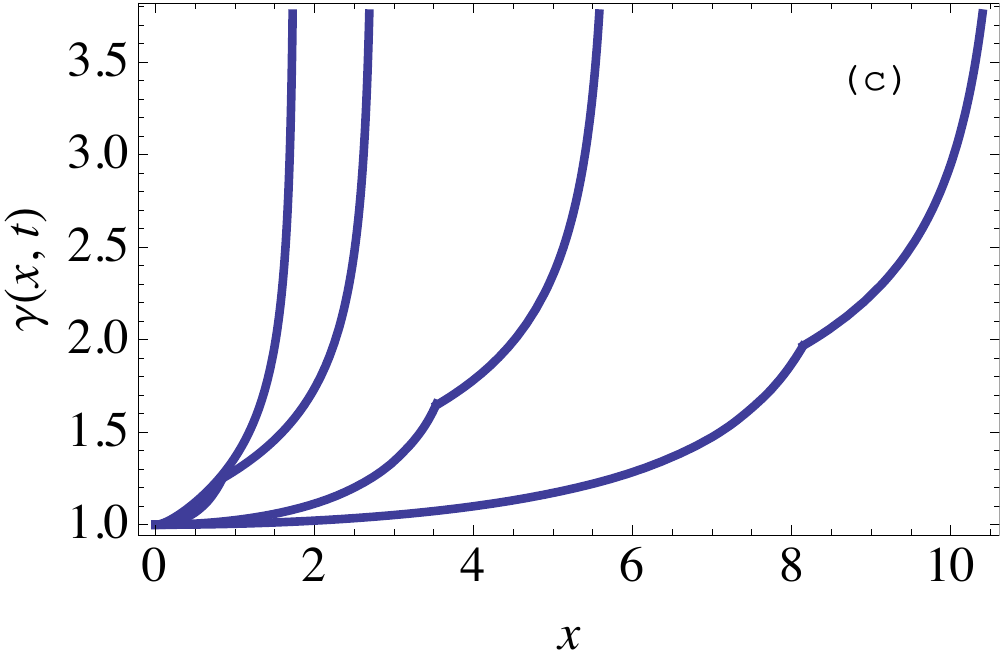}
\includegraphics[width=.49\linewidth]{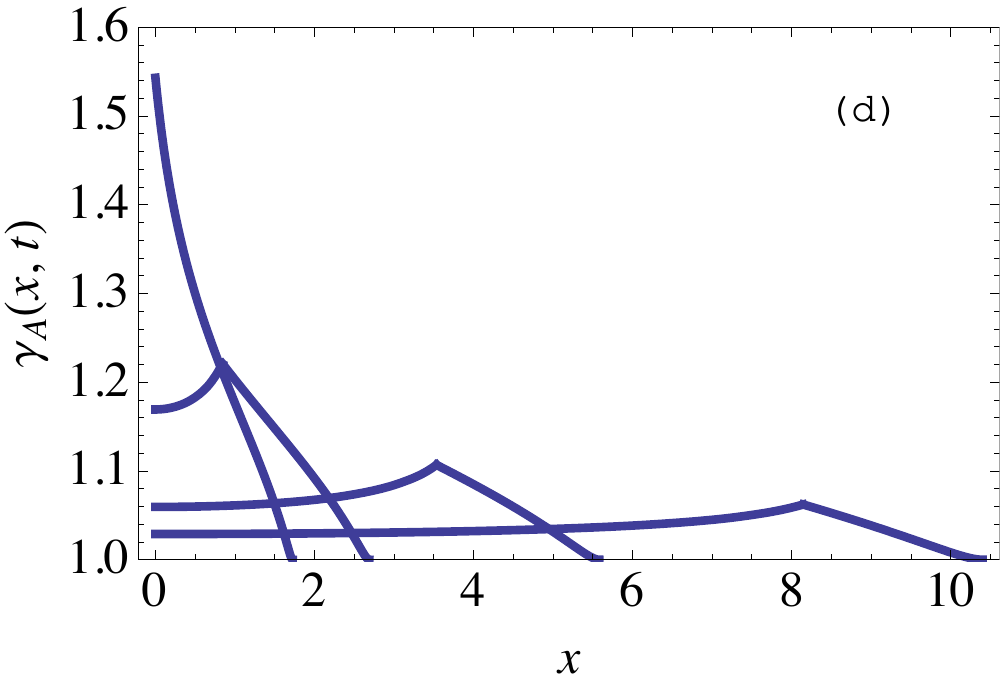}
\end{center}
\caption{Non-self-similar one-dimensional relativistic  expansion of a slab of magnetized plasma, initially  occupying $0< x<1$ and limited by a wall at $x=0$. Expansion proceeds in the positive direction.
Plots of (a) velocity and  (b) \Alfven speed, along with related \(\gamma\)-factors (c-d)
 with respect to position measured from the wall at \(x=0\) for times \(t=\)1, 2, 5, and 10.  Velocities are measured as fractional multiples of \(c\) and position, \(x\), is measured in the same units as time multiplied by the speed of light. Initially the plasma occupies the region \(0<x<\beta_{A,0}\) (Note: \(\beta_{A,0}=\tanh (1)\) in the present calculations)   and expands into the vacuum occupying \(x>\beta_{A,0}\) while an impenetrable wall is present at \(x=0\) (resulting in the reflection of the initial rarefaction wave at 
\(t=1\)).  The effect of finite distance from the wall occurs for times after \(t=1\) and can be observed as the discontinuities of the first derivatives in the velocities and \Alfven speeds, marking the separation between complex and simple regions of expansion.}
\label{fig:relplots}
\end{figure}

As a next step towards a numerical solution, \(J_2\) is calculated as a function of \(J_1\) and time by identifying the values of \(J_2\)
corresponding to a specified time through an interpolation method similar to the one used in the non-relativistic case, resulting in the calculation
of a function \(J_2\left(J_1,t\right)\). In order to determine the numerical solutions in the complex region for a given time, a collection of
points of the form \(\left( J_1, J_2\left(J_1;t\right), x\left[J_1,J_2\left(J_1;t\right)\right] \right)\), where we have emphasized only a parametric
dependence on time, are calculated based on a number of \(J_1\)values in the range \(\left(J_{1,x=0}(t),2\right)\).  This range represents the
\(J_1\) values in the complex region and \(J_{1,x=0}(t)\) represents the minimum value encountered at the point \(x=0\) subject to the condition
\(t=R\left[2\tanh ^{-1}\left(\beta _{A,0}\right),2\tanh ^{-1}\left(\beta _{A,0}\right);J_{1,x=0}(t),J_{1,x=0}(t)\right]\).  This condition
is based on the requirement that \(\beta (0,t)=0=\tanh \left(\frac{J_{1,x=0}-J_{2,x=0}}{2}\right)\) and hence \(J_{1,x=0}=J_{2,x=0}\).  Having
obtained a sufficient collection of points, the solutions
\ba && 
\beta (x,t)=\tanh \left[\frac{J_1(x,t) - J_2(x,t)}{2}\right]
\nn &&
\beta _A(x,t)=\tanh \left[\frac{J_1(x,t) + J_2(x,t)}{4}\right]
\ea
can be numerically calculated by interpolation. The functions \(\beta (x,t)\) and \(\beta _A(x,t)\) in the simple region of the expansion are known analytical functions of \(x\) and \(t\), more
specifically of a single variable \(\eta =\frac{x-x_0}{t}\), given in Ref. \cite{Simplewaves} as
\ba &&
\beta (\eta )=\frac{\delta _{A,0}{}^{4/3} \delta _{\eta }{}^{4/3}-1}{1+\delta _{A,0}{}^{4/3} \delta _{\eta }{}^{4/3}}
\nn &&
\beta _A(\eta )=1-2\frac{\delta _{\eta }{}^{2/3}}{\delta _{\eta }{}^{2/3}+\delta _{A,0}{}^{4/3}}
\label{relsimplereg}
\ea
The boundary, \(x_b(t)\), between the simple and complex regions of expansion is determined as a function of time in terms of \(\eta_b=\frac{x_b-x_0}{t}\) by the equation
\cite{Simplewaves}
\be
t=\left(\delta _{A,0}{}^2-1\right)\sqrt{\delta _{A,0}{}^4-1} \frac{1+\delta _{\eta_b}{}^2}{\left(\delta _{A,0}{}^{8/3}-\delta _{\eta_b}{}^{4/3}\right){}^{3/2}}
\ee
Combining the interpolated complex region solutions and the simple region solutions given in Eq.'s (\ref{relsimplereg}), complete solutions are plotted in Figure (\ref{fig:relplots}).

Previously, the problem of non-self-similar expansion of magnetized plasma was considered numerically in Ref.  \cite{2011MNRAS.411.1323G}.  Contrary to the initial claim in Ref.   \cite{2011MNRAS.411.1323G},  the presence of the wall is detrimental to acceleration, as can be seen from Fig. \ref{fig:relplots}.  Magnetic pressure-driven acceleration proceeds most efficiently while a given fluid element is causally disconnected from the wall, during the self-similar stage discussed in Ref.   \cite{Simplewaves}.

\section{Discussion}

In this paper we derived a number of analytical results  for  one-dimensional expansion of magnetized gas into plasma. First, we found the  self-similar expansion into vacuum of  a hot magnetized plasma. In this case the total pressure 
 has contribution  both form \Bf\ and from kinetic motion of particles. These two contribution  obey different equations of state, so effectively, we considered  relativistic self-similar expansion of a mixture of gasses with different adiabatic indices. 

Second, we derived relativistic hodograph and Darboux equations that describe arbitrary one-dimensional motion of magnetized plasma perpendicular to \Bf.  
 The  obtained resulting hodograph and Darboux  equations are  very powerful:  we reduced a  system of highly non-linear,  relativistic, time dependent equations describing  arbitrary (not necessarily self-similar)  dynamics of highly magnetized plasma to  a single {\it linear} differential equation.
 Using semi-analytical methods we calculated evolution of the flow parameters.

\bibliographystyle{apsrev}
\bibliography{/Users/maxim/Home/Research/BibTex} 

\appendix

\section{Non-Relativistic expansion of hot magnetized plasma into vacuum}

In the non-relativistic limit, the   equations of one-dimensional transverse  MHD read
\ba &&
\partial_t \rho + \partial_x ( v  \rho) =0
\nn &&
\rho (\partial_t v + v \partial_x v) = -\partial_x (B^2/2 +P)
\nn &&
\partial_t B + \partial_x ( B  \rho) =0
\nn &&
P = P_0 (\rho/\rho_0) ^\Gamma
\label{NR}
\ea
where $P$ is kinetic pressure, $B$ is \Bf\ divided by $\sqrt{4 \pi}$,  $\rho$ is density  is $v$ is plasma velocity. The kinetic pressure obeys a polytropic equations of state with index $\Gamma$.
As the initial condition,  we assume that at time $t=0$ plasma  occupies region $x<0$, while at  $x>0$ the   medium is  a vacuum. Initial homogeneous density, magnetic fields and kinetic pressures are $\rho_0$, $B_0$ and $P_0$ correspondingly. At time $t=0$ a barrier at $x=0$ is removed and the plasma starts expanding into vacuum while  a rarefaction wave propagates into the bulk plasma. In the initial state the \Alfven and sound velocities are
\ba &&
 v_{A,0}^2 = B_0^2 / \rho_0
 \nn &&
 c_{s,0}^2 = \Gamma P_0 /\rho_0
 \ea

 Let us assume that all the quantities depend on self-similar combination $\eta = z/t$. The system (\ref{NR}) then reduces to
\ba &&
(\eta -v) \rho_1 ' + v' \rho_1=0
\nn &&
(v-\eta)^2 = v_{A,0}^2 \rho_1 + c_{s,0}^2 \rho_1^{\Gamma -1} 
\label{31}
\ea
where  $\rho_1 = \rho/\rho_0$. 
 Eq. (\ref{31}) can be resolved for velocity $v(\rho_1)$,
 \be
 v= \eta \pm \sqrt{  v_{A,0}^2 \rho_1 + c_{s,0}^2  \rho_1^{\Gamma -1} } , 
 \label{v}
 \ee
 and an equation for $\rho_1$:
 \be
 \partial_\eta \rho_1 = \pm {  2 \sqrt{ v_{A,0}^2  + c_{s,0}^2  \rho_1^{\Gamma-2}} \rho_1^{3/2} \over
 3 v_{A,0}^2 + c_{s,0}^2  \rho_1^{\Gamma-2} (1+ \Gamma) }
\label{53}
 \ee
Eq. (\ref{53}) can be integrated
\be
\eta = C_1 \pm \left( {1+\Gamma \over \Gamma-1 } \sqrt{  v_{A,0}^2 \rho_1 + c_{s,0}^2  \rho_1^{\Gamma -1} }  -
2  {\Gamma -2 \over (3- \Gamma)(\Gamma-1) }{v_{A,0}^2 \over c_{s,0}}  \rho_1^{(3-\Gamma)/2} { _2 F _1}
\left({1\over 2}, {3-\Gamma \over 2 (2-\Gamma)}, { 7-3 \Gamma \over 2(2-\Gamma)}; - {  v_{A,0}^2 \over c_{s,0}^2}  \rho_1^{2-\Gamma}
\right)
\right)
\label{54}
\ee
where ${ _2 F _1}$ is a hypergeometric function.
Signs in Eqns (\ref{53}) and (\ref{54}) correspond to the choice in (\ref{v}).

The constant of integration $C_1$ can be found from the  condition that   at the front of the rarefaction wave, which propagates with fast velocity in the undisturbed medium $v_{f,0}$    and is located at $\eta_{RW} = -v_{f,0}= -  \sqrt{ v_{A,0}^2+ c_{s,0}^2} $, the plasma  density is undisturbed  $\rho_1 =1$:
\be
C_1 = {2 \over \Gamma -1} \sqrt{  v_{A,0}^2+ c_{s,0}^2}  \pm 2  {\Gamma -2 \over (3- \Gamma)(\Gamma-1) }{v_{A,0}^2 \over c_{s,0}}   { _2 F _1}
\left({1\over 2}, {3-\Gamma \over 2 (2-\Gamma)}, { 7-3 \Gamma \over 2(2-\Gamma)}; - {  v_{A,0}^2 \over c_{s,0}^2} 
\right)
\ee

\begin{figure}[h!]
 \begin{center}
\includegraphics[width=.49\linewidth]{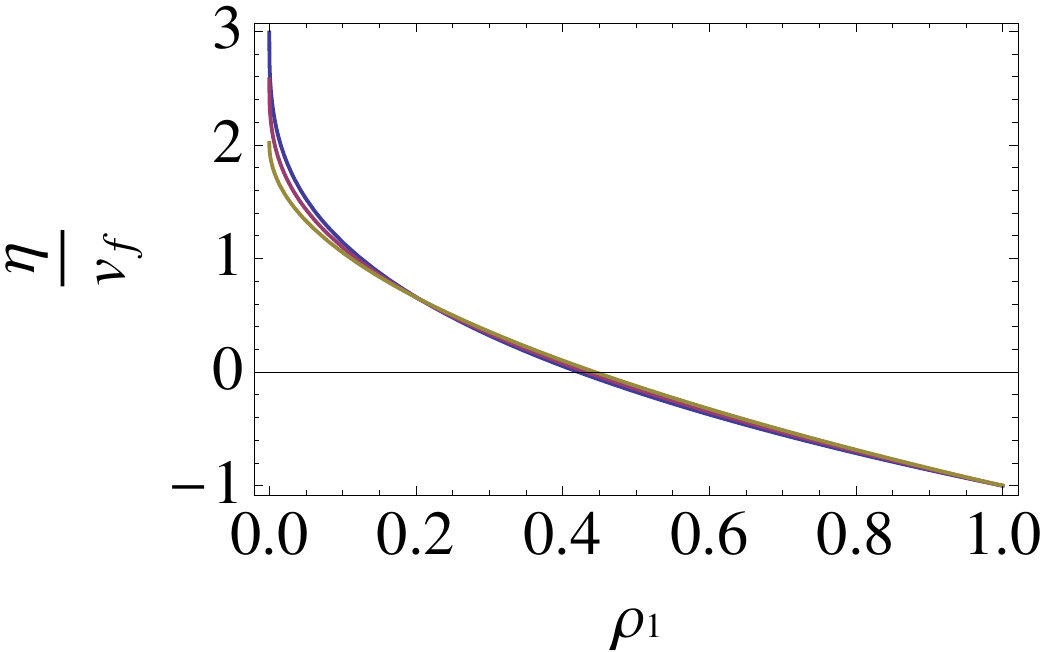}
\includegraphics[width=.49\linewidth]{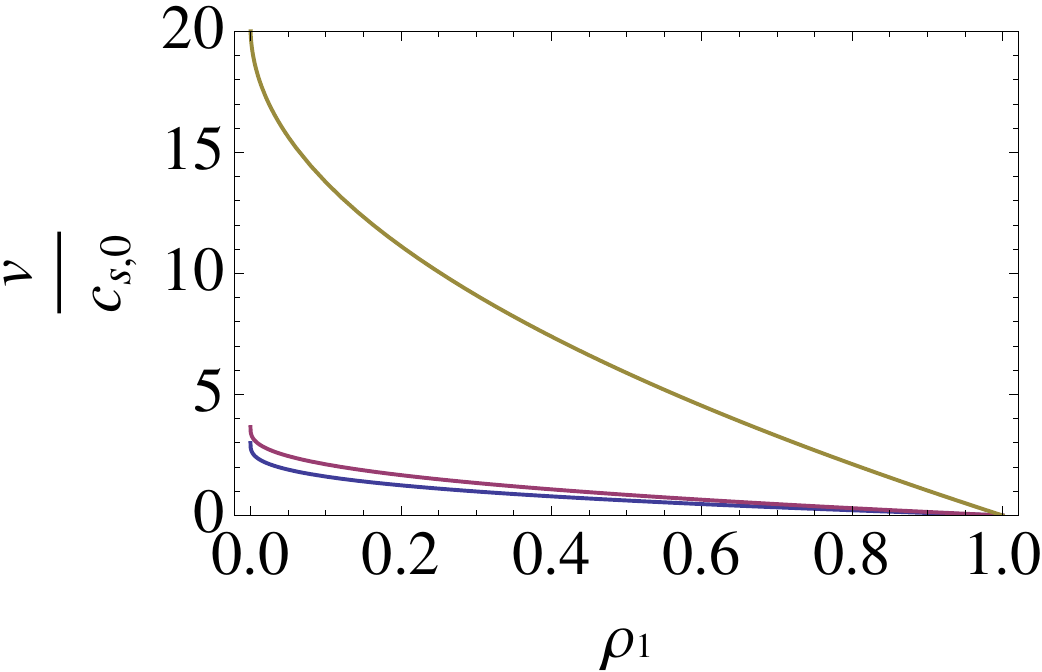}
\end{center}
\caption{Self-similar non-relativistic  expansion of magnetized  fluid into vacuum, Eq. (\ref{etaNR}), $\Gamma=5/3$.  {\it Left Panel}:   the value of the self-similar parameter  in terms of fast velocity, $\eta/v_{f,0}=\eta/\sqrt{c_{s,0}^2 +v_{A,0}^2} $ as a function of density $\rho_1=\rho/\rho_0$. 
{\it Right  Panel}: 
velocity in terms of sound velocity, $v/c_{s,0}$ as function of density. The front of the rarefaction wave is located at $\rho_1=1$, the vacuum interface is located at  $\rho_1=0$. Different curves correspond to  different values of $M_{A,0} =  v_{A,0}/ c_{s,0} = 0, \, 1, \,  10$ (top to bottom). 
}
\label{betalimitofeta}
\end{figure}

In particular, for adiabatic index $\Gamma=5/3$ the previous relations simplify
\ba &&
{ \eta  \over c_{s,0}} = \left(  2  (1+M_{A,0}^2)^{3/2} -  (2+3 M_{A,0}^2 \rho_1^{1/3}) \sqrt{1+M_{A,0}^2 \rho_1^{1/3}} \right) {1 \over M_{A,0}^2} 
\nn &&
v=2  { (1+M_{A,0}^2)^{3/2} -(1+M_{A,0}^2\rho_1^{1/3} )^{3/2}  \over M^2}  c_{s,0}
\nn &&
M_{A,0} =  {v_{A,0} \over c_{s,0}},
\label{etaNR}
\ea
see Fig. \ref{betalimitofeta}. 
The front of the rarefaction wave propagates with the fast speed and is located at  $\eta_{RW} = -  c_{s,0} \sqrt{1+M_{A,0}^2}$, while the vacuum interface, corresponding to $\rho_1=0$,    propagates with the velocity
\be
v_{\rm vac} = 2 {(1+M_{A,0}^2)^{3/2}  - 1 \over M^2} c_{s,0}
\ee

For unmagnetized plasma Eq. (\ref{etaNR}) gives 
\ba &&
\eta = (3-4 \rho_1^{1/3})  c_{s,0}
\nn &&
v_{\rm vac}  =3 c_{s,0},
\ea
while for cold  magnetized plasma
\ba &&
\eta = (2-2 \sqrt{\rho_1})  v_{A,0}
\nn &&
v_{\rm vac}  =2  v_{A,0}, 
\ea
in correspondence with the general solutions in media with adiabatic indices $\Gamma=5/3 $ and $\Gamma=2$ \cite{Stanyukovich,LLVI}.

More, generally, in case of unmagnetized plasma, $ v_{A,0}=0$,  and arbitrary adiabatic index $\Gamma$,
\be 
\eta =  \left( { 2   \over \Gamma -1} - {1+\Gamma \over \Gamma-1 } \rho_1 ^{ (\Gamma-1)/2}  \right) c_{s,0} .
\ee
(So that the vacuum interface expands with $ 2 c_{s,0}  /(\Gamma -1)$.)

The  Riemann invariants \cite{LLVI} are
\be
J_\pm = v \pm \int {d p \over \rho v_f} = v\pm 2 { c_{s,0} v_f^3 \over c_s^3 v_{A,0}^2}
\label{J}
\ee
where $p$ is total pressure, kinetic plus magnetic.  Relations (\ref{J}) assume that both sound and \Alfven velocities are non-zero.

The   Riemann invariants $J_\pm$ are constant along the corresponding characteristics $C_\pm$: $dz/dt= v \pm v_f$.
 We find then that the $C_-$ characteristics
are straight lines $x =\eta t$ , while $C_+$ characteristics  are determined by  
\be
{dz\over dt} = \left( 2{  (1+M^2)^{3/2} \over M^2} - {\sqrt{1+M^2 \rho_1^{1/3}} (2 +M^2 \rho_1^{1/3}) \over M^2}  \right) c_{s,0}
\ee
where $\rho_1(z/t) $ should be found from Eq. (\ref{etaNR}). This gives a transcendental equation for the characteristics:
\ba && 
\left( {dz\over dt} \right)^3 - \left( \eta + 4 { (1+M^2)^{3/2} \over M^2} \right)  \left( {dz\over dt} \right)^2+
{1\over 3} \left( \eta^2 + 8 { (1+M^2)^{3/2} \over M^2}  c_{s,0}   \eta+ {16 } \left( (1+M^2)^3 -1/9 \right) c_{s,0} ^2 \right)  {dz\over dt}  -
\nn &&
{1\over 27} \left(\eta^3 + 12 { (1+M^2)^{3/2} \over M^2}  c_{s,0}   \eta^2 +
16 {3 (1+M^2)^3+1\over M^4} c_{s,0}  ^2 \eta +
64 {\sqrt{1+M^2} ( 3 +6 M^2 +4 M^4 +M^2)\over M^4}  c_{s,0}  ^3 \right)
\ea

\section{Non-relativistic hodograph and Darboux equations for cold magnetized plasma}
\label{non-relattt}
Let us next  discuss non-relativistic one-dimensional  motion  of cold magnetized plasma. In this case the induction  and continuity   equations imply 
$B/\rho =$constant, so that  
the pressure, $P \propto B^2$ is related to density via adiabatic law $P \propto \rho^\Gamma$ with $\Gamma =2$. 

For polytropic index of  $\Gamma =2$, the corresponding  hodograph  equations becomes   \citep[][\S  105]{LLVI}
\ba &&
w \partial ^2 _w \chi - \partial ^2 _v \chi +  \partial  _v \chi=0 
\nn &&
t = {\partial \chi \over w}
\nn && 
x= v t -  {\partial \chi \over v}
\label{hodo1}
\ea
where $w=c_s^2=v_A^2$ is proper fluid enthalpy and $c_s=v_A$ is the sound (\Alfven) speed.

A general solutions of the equation (\ref{hodo1}) can be easily obtained. For example, a solution separable in $w,\beta$
is $\chi = e^{ \alpha \beta} I_0(2\alpha  \sqrt{w}) $. The main problem with solving the hodograph equation (\ref{hodo1}) (or the corresponding Darboux equation (\ref{Darboux})) in a simple application of one-dimensional expansion of gas into vacuum, is in finding solutions that satisfy boundary conditions, one given on the characteristics and another at a fixed velocity.

Equation (\ref{hodo1}) is often transformed taking Riemann invariants as   independent variables \citep{Stanyukovich}
\ba &&
J_1 = 2 c_s + v
\nn &&
J_2=2 c_s-v
\ea
$J_1$ is constant on the characteristics $dx/dt=v+c$, while $J_2$  is constant on the characteristics $dx/dt=v-c$.  In the initial state $J_1 =J_2 = 2 c_{s,0} = J_{1,0}=J_{2,0}$.

In terms of $\{J_1,J_2\}$, the hodograph equation becomes 
\be 
\partial_{J_1}\partial_{J_2} \chi +  {\cal N} {\partial_{J_1}\chi +\partial_{J_2} \chi \over J_1+J_2}=0, \, {\cal N} =1/2
\label{Darboux}
\ee
Eq. (\ref{Darboux}) is referred to as Darboux equation.  Darboux  equation played an important role in the development of fluid mechanics: analysis of the analogue of the Darboux equation lead Riemann to the formulation of the theory of hyperbolic equations.

For integer values of the coefficient ${\cal N}=(1/2) (3 - \Gamma)/(\Gamma-1)$,  Darboux equation can be reduced to a one-dimensional wave equation. For the case of interest, $\Gamma=2$,  ${\cal N}=1/2$, this is not possible.


Equivalently, using constancy of the Riemann invariants on the characteristics, one find
\ba &&
\partial_{J_2} x ={ 3 J_1 - J_2 \over 4} \partial_{J_2} t
\nn &&
\partial_{J_2} x ={  J_1 - 3 J_2 \over 4} \partial_{J_2} t
\label{xt}
\ea
On can then write Darboux equation for time variable:
\be 
\partial_{J_1,J_2}t +  {3\over 2} {\partial_{J_1}t \partial_{J_2} t \over J_1+J_2}=0.
\label{DarbouxTime}
\ee


The main mathematical difficulty in solving  Darboux equation (\ref{Darboux})  with boundary conditions (\ref{boundary}) is that one boundary condition is given on the characteristics, while the other at a fixed values of velocity.  For integer values of ${\cal N}$ \cite{LLVI} give the solution; \eg\  for ${\cal N}=1$, $\chi = (J_2^2 - J_{2,0}^2)/(J_1+J_2)$. 

A general solutions of the Darboux equation (\ref{Darboux})   can be expressed in terms of the corresponding Riemann function (the analogue of the Green's function)  \citep[][Eq.3.71]{1967npde.book.....A}
\be 
B(J_{1,0},J_{2,0}, J_1,J_2) = \left(J_1+J_2 \over J_{1,0} + J_{2,0}\right)^{\cal N} {_2 F_1} \left(1-{\cal N}, {\cal N}, 1; - {(J_{1,0}-J_1)(J_{2,0}-J_{2}) \over (J_1+J_2 )( J_{1,0} + J_{2,0})} \right) 
\ee
It turns out that the corresponding Darboux equation  for time, Eq. (\ref{DarbouxTime}) with boundary condition $t=t_0$, the moment of reflection,  when $J_1=J_2 = 2$  can be solved explicitly \citep[see][Eq. 82.17]{1948sfsw.book.....C}
\be 
t=t_0 B(J_{1,0},J_{2,0}, J_1,J_2) 
\ee
(In passing we note that the equations of the one-dimensional fluid motion with adiabatic index $\Gamma=2$ are equivalent to shallow water equation. The corresponding problem of a dam break with a finite lock length has been solve in Ref. \cite{2006JFM...569...61H}.)

This gives
\be
t(J_1,J_2)= { 64 \over \sqrt{J_{1} + J_{1,0}} \sqrt{J_{2} + J_{2,0}} } {_2F_1} \left( {3\over 2} , {3\over 2},1, { (J_{1,0}- J_{1} ) ( J_{2,0} - J_2) \over (J_{1,0}+ J_{1} ) ( J_{2,0} + J_2)}\right)
\label{t}
\ee
where $_2F_1$ is the hypergeometric function. 

Coordinate $x$ then can then be derived from Eq. (\ref{xt}):
\be
x= {3 J_1 +J_2 \over 4} t   - {1\over 4} \int_{J_{2,0}}^{J_2} t d J_2
\label{xx}
\ee


In the numerical solution of the non-relativistic case, the initial conditions are such that the velocity, \(v(x,0)=0\), the Alfv{\' e}n
speed, \(c_A(x,0)=1\), the plasma occupies the region \(0<x<1\), and an impenetrable wall is present at \(x=0\).  These conditions are exactly
analogous to those given in Ref. \cite{2006JFM...569...61H} for the case  of  large Froud number.   The front of the expansion of plasma propagates as 
$ x_{\text{Front}}(t)=2t+1
$, 
corresponding to the foremost \(J_2\) 
characteristic, 
with Riemann invariant value \(J_2=-2\), emanating from \(x=1\) at \(t=0\). (We refer to characteristics for which \(\frac{dx}{dt}=v-c\)as \(J_2\) characteristics and characteristics for which \(\frac{dx}{dt}=v+c\)
as \(J_1\) characteristics, the names thus referring to the Riemann invariant which is constant on the specified characteristic)  The linear relationship governing the expansion front position as a function of time is obtained by integration of (Eq. 48) .  At time \(t=1\text{}\),
the initial rarefaction wave reaches the wall at \(x=0\) and is reflected.  Boundary conditions are such that plasma velocity is 0 at the wall location
x=0 (i.e., \(J_1-J_2=0\) when \(x=0\)). The constancy of the Riemann invariant \(J_2=2c-v\text{  }\)on the characteristic \(\frac{dx}{dt}=v-c\), implies
the leading rearward propagating characteristic emanating from \(x=1\) has value \(J_2=2\) on the path \(x=1-t\).  Correspondingly, the \(J_1\) characteristic
emanating from \(x=1\) at \(t=0\) has the value 2, satisfying the condition  \(J_1-J_2=0\) when \(x=0\). All  \(J_1\) characteristics will have
a constant value \(J_1=2\) for \(t<1\) in the region \(\{1-t<x<2t+1\}\), satisfying the initial conditions of \(v(x,0)=0\) and \(c_A(x,0)=1\). {
}For \(t>1,\) the \(J_1\) characteristics begin to influence the flow in the region \(0<x<x_b(t)\).  \(\text{}\)The discontinuities in the first
derivatives of the plots after \(t=1\) separate the simple and complex region of the flow. The conditions determined by eq.'s \ref{xt} in the simple region where  \(J_1=2\) are
\ba &&
\partial _{J_2}x=\frac{1}{4}\left(6-J_2\right)\partial _{J_2}t
\nn &&
x=1+\frac{1}{4}\left(2-3J_2\right)t
\label{simplecond}
\ea
We can numerically solve for the position, \(x_b(t)\), of the boundary between the simple and complex regions. This point is given by the implicit equation
\be
x_b(t)=1+2\cdot \frac{2-3 J_2\left(x_b,t\right)}{\left(2+J_2\left(x_b,t\right)\right){}^{3/2}}
\ee
 The solution for \(J_2\) (and therefore also velocity and Alfv{\' e}n speed) in the simple region,where \(J_1=2\), can also be obtained
from Eq. (\ref{simplecond}), which gives
\be 
J_2(x,t)=\frac{1}{3}\left(2-4\left( \frac{x-1}{t}\right)\right)
\ee
The functions \(\left\{t\left(J_1,J_2\right) ,x\left(J_1,J_2\right)\right\}\) are given, in the complex region, by Eq.{'}s
(\ref{t}) and (\ref{xx}). The first step in numerically constructing the functions  \(\left\{J_1(x,t),J_2(x,t)\right\}\) in the complex region is accomplished
by numerically constructing a function \(J_2\left(J_1,t\right)\) by interpolating values of \(t\left(J_1,J_2\right)\) at numerous values of \(J_2\)
over the range \(\left[-2,J_1\right]\) (this range covers all possible values of  \(J_2\) in the complex region).  Having achieved computationally
constructing a  function \(J_2\left(J_1,t\right)\), it is then possible to calculate position as a function of time and \(J_1\): \(x\left(J_1,t\right)=x\left(J_1,J_2\left(J_1,t\right)\right)\). A procedure similar to that used in calculating  \(J_2\left(J_1,t\right)\) from \(t\left(J_1,J_2\right)\) is then be employed to calculate
 \(J_1(x,t)\) and thereby \(J_2(x,t)= J_2\left(J_1(x,t), t\right)\). The procedure described allows for the Riemann invariants \(J_{1,2}\),
and hence \(v(x,t) \) and \(c_A(x,t)\) via Eq. (\ref{46}), to be calculated in the complex region.

   \begin{figure}[h!]
 \begin{center}
\includegraphics[width=.49\linewidth]{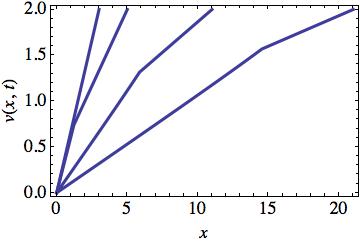}
\includegraphics[width=.49\linewidth]{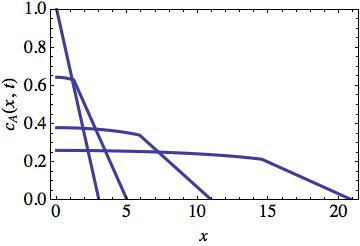}
\end{center}
\caption{Non-relativistic expansion of magnetized plasma limited by the wall at $x=0$. Plots of the plasma velocity, \(v(x,t)\), and Alfven speed, \(c(x,t)\), as functions of distance from the wall at \(x=0\) for one-dimensional flow satisfying Eq.'s (\ref{hodo1}), (\ref{boundary}). Initial conditions are such that gas is initially at rest occupying the region \(0<x<1\) with Alfven speed \(c_A=1\). An impenetrable wall is present at \(x=0\) and the gas expands into vacuum at \(t=0\).  Times shown in this plot are: \(t=1,2,5, \text{and } 10\).}
\label{nonrelplots}
\end{figure}

\end{document}